\documentclass[12pt]{article}
\usepackage{epsfig,amsfonts,amsthm}
\usepackage{amsmath,amssymb}
\usepackage{array}
\usepackage{comment}
\usepackage{hyperref}
\usepackage{cite}
\usepackage{color}
\usepackage{pb-diagram}
\usepackage{slashed}
\usepackage{cancel}
\newcommand{\be}{\begin{equation}}
\newcommand{\ee}{\end{equation}}
\newcommand{\bea}{\begin{eqnarray}}
\newcommand{\eea}{\end{eqnarray}}

\usepackage{float}
\usepackage[justification=centering]{caption}
\usepackage{tikz}
\usetikzlibrary{decorations.pathmorphing,decorations.markings}
\usepackage{graphicx}
\tikzset{
photon/.style={decorate, decoration={snake,amplitude=2pt, segment length=5pt}, draw=black},
particle/.style={draw=black, postaction={decorate}, decoration={markings,mark=at position .5 with {\arrow[draw=black]{>}}}},
antiparticle/.style={draw=black, postaction={decorate}, decoration={markings,mark=at position .5 with {\arrow[draw=black]{>}}}},
gluon/.style={decorate, draw=black, decoration={coil,amplitude=4pt, segment length=5pt}}
goldstone/.style={draw=green,postaction={decorate},decoration={markings,mark=at position .5 with {\arrow[draw=blue]{>}}}}
}

\newcommand{\doublet}[2]{ \left( \begin{array}{c}#1 \\ #2 \end{array}\right) }

\usepackage{color}
\definecolor{myorange}{rgb}{1,0.5,0}

\begin{document}
\bibliographystyle{OurBibTeX}

\title{\hfill ~\\[-30mm]
\begin{footnotesize}
\hspace{80mm}
HIP-2021-6/TH \\
\end{footnotesize}
\vspace{5mm}
\textbf{CP-violating inflation\\
and\\
its cosmological imprints}
}
\date{}

\author{\\[-5mm]
Venus Keus\footnote{E-mail: {\tt venus.keus@helsinki.fi}} $^{1,2}$,
Kimmo Tuominen\footnote{E-mail: {\tt kimmo.i.tuominen@helsinki.fi}} $^{1}$
\\ \\
\emph{\small  $^1$ Department of Physics, University of Helsinki,}\\
\emph{\small P.O.Box 64, FI-00014 Helsinki, Finland Helsinki, Finland}\\
\emph{\small $^2$ School of Physics and Astronomy, University of Southampton,}\\
\emph{\small Southampton, SO17 1BJ, United Kingdom}
\\[4mm]}

\maketitle

\begin{abstract}
\noindent
{We study models with several $SU(2)$ scalar doublets where the inert doublets
have a non-minimal coupling to gravity and play the role of the inflaton.
We allow for this coupling to be complex, thereby introducing CP-violation
- a necessary source of the baryon asymmetry - in the Higgs--inflaton couplings.
We investigate the inflationary dynamics of the model and discuss how the
CP-violation of the model is imprinted on the particle asymmetries
after inflation in the hot big bang universe.
}
 \end{abstract}
\thispagestyle{empty}
\vfill
\newpage
\setcounter{page}{1}

\section{Introduction}
\label{sec:intro}
The Standard Model (SM) of particle physics has been extensively tested and is in great agreement with experimental data, with
its last missing particle – the Higgs boson – discovered by ATLAS and CMS experiments at the CERN Large Hadron Collider (LHC) \cite{Aad:2012tfa,Chatrchyan:2012ufa}.
Although the properties of the observed scalar are in agreement with those of the SM-Higgs boson, it may just be one member of an extended scalar sector.
Even though so far no signs of
new physics have been detected, it is well understood that the SM of particle physics is incomplete.

Cosmological and astrophysical observations imply a large dark matter (DM) component
in the energy budget of the universe.
Within the particle physics setting, this would be a particle which is stable on cosmological time scales, cold, non-baryonic, neutral and weakly interacting \cite{Ade:2015xua}.
A particle with such characteristics does not exist in the SM.
Another shortcoming of the SM is the lack of an explanation for the origin of the observed matter-antimatter asymmetry in the universe.
One of the most promising baryogenesis scenarios is electroweak baryogenesis (EWBG) \cite{Morrissey:2012db}, which produces the baryon excess during the electroweak phase transition (EWPT). Although the SM in principle contains all required ingredients for EWBG, it is unable to explain the observed baryon excess due to its insufficient amount of CP-violation \cite{Gavela:1993ts,Huet:1994jb,Gavela:1994dt} and the lack of a first-order phase transition \cite{Kajantie:1995kf}.

Furthermore, in its current form, the SM fails to incorporate cosmic inflation
in a satisfactory manner.
Inflation is a well-motivated theory predicting a period of exponential expansion
in the early universe
which explains the generation of primordial density fluctuations seeding structure formation,
flatness, homogeneity and isotropy of the universe
\cite{Hawking:1982cz,Starobinsky:1982ee,Sasaki:1986hm,Mukhanov:1988jd}.
The simplest models of inflation in best agreement with observations are those driven by a
scalar field, the \textit{inflaton}, with a standard kinetic term, slowly rolling down its
smooth potential.
At the end of inflation, the inflaton which naturally is assumed to have couplings with the SM-
Higgs, dumps its energy into the SM bath during the \textit{reheating} process which populates
the universe with SM particles.

Scalars with non-minimal couplings to gravity
%which are light during inflation
are well-motivated inflaton candidates since they acquire fluctuations proportional to the inflationary scale and can drive the inflation process in the early universe,
as in the Higgs-inflation model \cite{Bezrukov:2007ep} where
the SM-Higgs plays the role of the inflaton, and $s$-inflation models \cite{Lerner:2009xg,Enqvist:2014zqa} where the SM is extended by a singlet scalar.
Extensive studies have been carried out in simple one singlet or one doublet scalar extensions of the SM (see e.g. \cite{Gong:2012ri,Choubey:2017hsq,Englert:2011yb,Branco:2011iw} and references therein).
These models, however, by construction can only partly provide a solution to the
main drawbacks of the SM.
For example, to incorporate both CP-violation and DM into the model
one has to go beyond simple scalar extensions of the SM \cite{Keus:2016orl};
see also e.g. \cite{Cordero-Cid:2016krd, Cordero:2017owj,Cordero-Cid:2018man,Cordero-Cid:2020yba,Keus:2020ooy,Keus:2019szx}.

It is therefore theoretically appealing to have a more coherent setting where
different motivations of beyond SM (BSM) frameworks could be simultaneously investigated.
For example, in non-minimal Higgs frameworks with conserved discrete symmetries one
can accommodate stabilised DM candidates. Moreover, the extended scalar potential
could provide new sources of CP-violation and accommodate a strong first order
phase transition \cite{Ahriche:2015mea}. Collider searches can constrain these model frameworks by
excluding or discovering the existence of the spectrum of new states.

In this paper we introduce a model where a source of CP-violation
originates from the couplings of the inflation.
Through the process of reheating this is transmitted to an asymmetry within the
SM and can furthermore seed
the generation of an excess of matter over antimatter during the
evolution of the early universe.
We describe these dynamics in the context of a $Z_2$ symmetric
3-Higgs Doublet Model (3HDM) with a CP-violating extended dark sector,
which also provides a viable DM candidate, new sources of CP-violation and
a strong first-order EWPT \cite{Cordero-Cid:2016krd,Keus:2016orl,Cordero:2017owj,Cordero-Cid:2018man,Cordero-Cid:2020yba,Keus:2020ooy}.
We study the inflationary dynamics of this set-up and outline its main
consequences. In a future work we aim to continue to complement this study
by more thorough analysis of EWBG and DM observables as well as a phenomenological analysis
towards LHC searches for new physics.

The paper is organized as follows.
In Section \ref{sec:potential} we present the scalar potential and explore the inflationary
dynamics. In Section \ref{sec:slow-roll}, we discuss the inflationary imprints of our novel CP
violating inflation phenomena. In Section \ref{sec:reheating}, we discuss the inflaton decay into the SM particles and possible consequences.
In Section \ref{sec:conclusions} we draw our conclusions and discuss the outlook for further
work.

\section{The scalar potential}
\label{sec:potential}
\subsection{General definitions}
A 3HDM scalar potential which is symmetric under a group $G$ of phase rotations, can be written as the sum of two parts: $V_0$ with terms symmetric under any phase rotation, and $V_G$ with terms symmetric under $G$ \cite{Ivanov:2011ae,Keus:2013hya}. As a result, a $Z_2$-symmetric 3HDM can be written as\footnote{We ignore additional $Z_2$-symmetric terms that can be added to the potential, e.g.,
$
(\phi_3^\dagger\phi_1)(\phi_2^\dagger\phi_3),
$
$
(\phi_1^\dagger\phi_2)(\phi_3^\dagger\phi_3),
$
$
(\phi_1^\dagger\phi_2)(\phi_1^\dagger\phi_1)
$
and
$
(\phi_1^\dagger\phi_2)(\phi_2^\dagger\phi_2),
$
as they do not change the phenomenology of the model \cite{Cordero-Cid:2018man}.}:
\bea
\label{eq:V0-3HDM}
V&=&V_0+V_{Z_2}, \\
V_0 &=& - \mu^2_{1} (\phi_1^\dagger \phi_1) -\mu^2_2 (\phi_2^\dagger \phi_2) - \mu^2_3(\phi_3^\dagger \phi_3) \nonumber\\
&&+ \lambda_{11} (\phi_1^\dagger \phi_1)^2+ \lambda_{22} (\phi_2^\dagger \phi_2)^2  + \lambda_{33} (\phi_3^\dagger \phi_3)^2 \nonumber\\
&& + \lambda_{12}  (\phi_1^\dagger \phi_1)(\phi_2^\dagger \phi_2)
 + \lambda_{23}  (\phi_2^\dagger \phi_2)(\phi_3^\dagger \phi_3) + \lambda_{31} (\phi_3^\dagger \phi_3)(\phi_1^\dagger \phi_1) \nonumber\\
&& + \lambda'_{12} (\phi_1^\dagger \phi_2)(\phi_2^\dagger \phi_1)
 + \lambda'_{23} (\phi_2^\dagger \phi_3)(\phi_3^\dagger \phi_2) + \lambda'_{31} (\phi_3^\dagger \phi_1)(\phi_1^\dagger \phi_3),  \nonumber\\
 V_{Z_2} &=& -\mu^2_{12}(\phi_1^\dagger\phi_2)+  \lambda_{1}(\phi_1^\dagger\phi_2)^2 + \lambda_2(\phi_2^\dagger\phi_3)^2 + \lambda_3(\phi_3^\dagger\phi_1)^2  + h.c. \nonumber
\eea
where the three Higgs doublets, $\phi_{1},\phi_2,\phi_3$, transform under the $Z_2$ group, respectively, as
\be
\label{eq:generator}
g_{Z_2}=  \mathrm{\rm diag}\left(-1, -1, +1 \right).
\ee
The parameters of the $V_0$ part of the potential are real by construction. We allow for the parameters of $V_{Z_2}$ to be complex, using the following notation throughout the paper
\be
\label{eq:complex-params}
\lambda_{j} = |\lambda_{j}| \, e^{i\, \theta_{j}} \quad (j = 1,2,3), \quad
\quad
\mbox{and}
\quad
\mu^2_{12} = |\mu^2_{12}| \, e^{i\, \theta_{12}}\,.
\ee
The composition of the doublets is as follows:
\be
\phi_1= \doublet{$\begin{scriptsize}$ H^+_1 $\end{scriptsize}$}{\frac{H_1+iA_1}{\sqrt{2}}},\quad
\phi_2= \doublet{$\begin{scriptsize}$ H^+_2 $\end{scriptsize}$}{\frac{H_2+iA_2}{\sqrt{2}}}, \quad
\phi_3= \doublet{$\begin{scriptsize}$ G^+ $\end{scriptsize}$}{\frac{v+h+iG^0}{\sqrt{2}}},
\label{explicit-fields}
\ee
where $\phi_1$ and $\phi_2$ are the $Z_2$-odd \textit{inert} doublets,
$\langle \phi_1 \rangle = \langle \phi_2 \rangle =0$, and $\phi_3$ is the one
$Z_2$-even \textit{active} doublet, which at low energy attains a vacuum expectation
value (VEV)
$\langle \phi_3 \rangle =v/$\begin{scriptsize}$ \sqrt{2} $\end{scriptsize} $ \neq 0$.
The doublet $\phi_3$ plays the role of the SM Higgs doublet, with $h$ being the
SM Higgs boson and $G^\pm,~ G^0$ the would-be Goldstone bosons.
Note that
according to the $Z_2$ generator in Eq.~\eqref{eq:generator}
the symmetry of the potential is respected by the vacuum
$(0,0,v/$\begin{scriptsize}$ \sqrt{2} $\end{scriptsize}$)$. In this paper we consider the scenario where
the components of the inert doublets act as inflation candidates and reheat the
universe at the end of inflation through their interactions with the SM-Higgs and
gauge bosons. Note that at the scales relevant for inflation we can take the VEV
of the active doublet to be zero, $\langle \phi_3 \rangle = 0$.

Furthermore, CP-violation is only introduced in the \textit{inert} sector which is forbidden
from mixing with the \textit{active} sector by the conservation of the $Z_2$ symmetry. As a
result, the amount of CP-violation is not limited by electric dipole moments \cite{Cordero-Cid:2016krd}.
The lightest particle amongst the CP-mixed neutral fields from the inert doublets is a
viable DM candidate and stable due to the unbroken $Z_2$ symmetry.
In this paper, we focus on the inflationary dynamics of the model and shall not discuss DM
implications of the model any further.

\subsection{Potential for the inflaton}

We start by rewriting the doublets in the unitary gauge and ignore the charged scalars
(since they do not affect the inflationary dynamics).
\be
\phi_1= \frac{1}{\sqrt{2}} \left(\begin{array}{c}
0 \\[2mm]
h_1 +i \eta_1
\end{array} \right)
,\quad
\phi_2= \frac{1}{\sqrt{2}} \left(\begin{array}{c}
0 \\[2mm]
h_2 + i \eta_2
\end{array} \right)
, \quad
\phi_3= \frac{1}{\sqrt{2}} \left(\begin{array}{c}
0 \\[2mm]
h_3
\end{array} \right).
\label{eq:explicit-fields}
\ee

The action of the model in the Jordan frame is
\bea
\label{Eq:action-Jordan}
S_J &=& \int d^4x \sqrt{-g}
\bigg[  -\frac{1}{2} M^2_{pl} R - D_\mu \phi_1^\dagger  D^\mu \phi_1
- D_\mu \phi_2^\dagger  D^\mu \phi_2
- D_\mu \phi_3^\dagger  D^\mu \phi_3
\\
&& \hspace{2.3cm}
- V(\phi_1, \phi_2,\phi_3)
- \biggl(\xi_1 |\phi_1|^2  +\xi_2 |\phi_2|^2 +\xi_3 |\phi_3|^2
+ \xi_4 (\phi^\dagger_1 \phi_2)  +\xi_4^* (\phi^\dagger_2 \phi_1 ) \biggr)R
\bigg],
\nonumber
\eea
where $R$ is the Ricci scalar, $M_{pl}$ is the reduced Planck mass and the parameters $\xi_i$ are dimensionless couplings of the scalar doublets to gravity.
%(quantum effects invariably give rise to such couplings at Planck scales).
Note that, in principle, $\xi_4$ could be a complex parameter for which we use the notation
\be
\label{eq:xi4-phase}
\xi_4 = |\xi_4|\, e^{i\theta_4}\,.
\ee

In Eq.~\eqref{Eq:action-Jordan} the covariant derivative, $D_\mu$, contains couplings of the scalars with the gauge bosons. However, for the dynamics during the inflation,
%during inflation, there are no fields other than the inflaton and
the covariant derivative is reduced to the normal derivative
$D_\mu\rightarrow \partial_\mu$. The minus sign in the kinetic terms follows the metric convention of $(-,+,+,+)$.

Since we identify the two inert doublets with inflaton, we assume that the energy density of
$\phi_3$ is sub-dominant during inflation. Therefore, the
%relevant part of the potential \blue{(or:
part of the potential relevant for inflation
is
\bea
\label{eq:approxscalarpot}
V &=& - \mu^2_{1} (\phi_1^\dagger \phi_1) -\mu^2_2 (\phi_2^\dagger \phi_2)
+ \lambda_{11} (\phi_1^\dagger \phi_1)^2+ \lambda_{22} (\phi_2^\dagger \phi_2)^2 \\
&& + \lambda_{12}  (\phi_1^\dagger \phi_1)(\phi_2^\dagger \phi_2)
 + \lambda'_{12} (\phi_1^\dagger \phi_2)(\phi_2^\dagger \phi_1)
 -\mu^2_{12}(\phi_1^\dagger\phi_2)+  \lambda_{1}(\phi_1^\dagger\phi_2)^2 + h.c. \nonumber
\eea
Due to local SU(2) invariance, we can rotate away one of the CP-odd fields,
say $\eta_2$.
%due to the re-phasing invariance.
Such a transformation is equivalent to taking the $\eta_2 \to 0$ limit, and
we assume this limit to be taken when writing the fields in terms
of components in Eq.~\eqref{eq:explicit-fields}.

%Furthermore, we assume that $\xi_4\gg \xi_1,\xi_2,\xi_3$ and retain only the
%contribution from the nonminimal coupling terms proportional to $\xi_4$.
% For the inert doublets to serve as inflaton candidates, one needs to ensure that
% \be
% \frac{\lambda_{11}}{\xi_1^2} , \, \frac{\lambda_{22}}{\xi_2^2}   \ll \frac{\lambda_{33}}{\xi_3^2}
% \ee
% This condition can be easily satisfied when the self-coupling parameter of the active doublet $\lambda_{33}$ is of the same order of its coupling to gravity, i.e. $\lambda_{33}\sim \xi_3$,  while the relevant couplings of the two inert doublets $\lambda_{11}, \lambda_{22}$ are of order 1 and much smaller than their non-minimal coupling to gravity, $\lambda_{11}, \lambda_{22} \sim 1\ll\xi_1,\xi_2$.
% To focus on the effect of the CP-violating parameter $\xi_4$, we assume $\xi_4 \gg \xi_1,\xi_2$.

To facilitate the analysis, we apply a conformal transformation from %the physical or
the Jordan frame, which contains terms with scalar-gravity quadratic couplings,
to the Einstein frame with no explicit couplings to gravity \cite{Kaiser:2010ps}.
Physical observables are invariant under this frame transformation. The two frames
are equivalent after the end of inflation when the transformation parameter equals unity.

The action in the Einstein frame can be written as
\be
S_E=\int d^4x \sqrt{-\tilde g}\left[-\frac{1}{2}M_{pl}^2\tilde R-\frac{1}{2}\tilde{g}^{\mu\nu}\, G_{ij} \,\partial_\mu \varphi_i \partial_\nu\varphi_j-\tilde{V}\right],
\label{eq:action-Einstein}
\ee
where $\tilde{V} = V/\Omega^4$ is the potential in the Einstein frame following
the conformal transformation
\bea
\tilde{g}_{\mu\nu}&=&
\Omega^2 g_{\mu\nu},
\nonumber\\
G_{ij}&=&
\frac{1}{\Omega^2}\delta_{ij}+\frac{3}{2}\frac{M_{pl}^2}{\Omega^4}\frac{\partial\,\Omega^2}{\partial\,\varphi_i}\frac{\partial\,\Omega^2}{\partial\,\varphi_j},
\label{eq:conformal-transf}
\eea
where $\varphi_k=h_{1},h_{2},\eta_{1}$, and the transformation parameter
\be
\label{eq:def-Omega}
\Omega^2 =
%1+\frac{1}{M_{pl}^2} \biggl[
%\xi_1 (h_1^2+\eta_1^2) + \xi_2 (h_2^2+\eta_2^2) +\xi_3 h_3^2
%+ \xi_4\left(h_1h_2+\eta_1 \eta_2 + i (h_1\eta_2 - h_2 \eta_1) \right)\biggr]
% + h.c.
%\nonumber
%\ee
%which we approximate as ($\xi_4 \gg \xi_1,\xi_2,\xi_3$)
%\be
%\label{eq:approx-xi4}
%\Omega^2 \approx
1 +\frac{\xi_1}{M_{pl}^2}(h_1^2+\eta_1^2)+\frac{\xi_2}{M_{pl}^2}h_2^2+\frac{2 |\xi_4|}{M_{pl}^2}
\biggl( h_1h_2 c_{\theta_4} +\eta_1 h_2 s_{\theta_4}\biggr)
\ee
using the shorthand notation $c_{\theta_k} = \cos\theta_k $ and $s_{\theta_k} = \sin\theta_k $ throughout the paper.

The prefactor $G_{ij}$ in Eq.~\eqref{eq:conformal-transf} leads to mixed kinetic terms.
We introduce the reparametrisation
\be
\label{eq:def-A}
A =\sqrt{\frac{3}{2}} \, M_{pl} \, \log (\Omega^2)
\qquad \mbox{with} \qquad
\frac{\partial\,\Omega^2}{\partial\,\varphi_k} =
\sqrt{\frac{2}{3}}\, \frac{\Omega^2}{M_{pl}} \, \frac{dA}{d\varphi_k}
\ee
which reduces the kinetic terms to the diagonal form
%\blue{
\be
\label{eq:diag-kinetic}
\tilde{g}_{\mu\nu} \, G_{ij} \,  \partial_\mu \varphi_i \partial_\nu\varphi_j
=
\Omega^2 g_{\mu\nu} \biggl( \frac{\delta_{ij}}{\Omega^2}
+  \frac{\partial A}{\partial \varphi_i}  \, \frac{\partial A}{\partial \varphi_j} \biggr)
\partial_\mu \varphi_i \partial_\nu \varphi_j
\, = \,
\partial_\mu \varphi_i \partial_\mu \varphi_i
+ \Omega^2 \, \partial_\mu A \,  \partial_\mu A
\ee
%}
%Note that $A$ is a complex field through CP-violating contributions from $\Omega^2$.

%\subsection{The potential in the Einstein frame}

To write the potential in the Einstein frame, %we note that
%all scalar fields are present in the lagrangian but during inflation, fields other than the %inert doublets components have negligible contribution.
%Therefore,
we keep only terms in the potential in Eq.~\eqref{eq:approxscalarpot}
which are quartic in $h_{1,2}$ and $\eta_{1}$.
This reduces the potential to
\bea
\tilde{V} &\approx & \frac{1}{4 \, \Omega^4} \biggl[
\lambda_{11} (h_1^2 + \eta_1^2)^2
+\lambda_{22}  h_2^4
+(\lambda_{12}+\lambda'_{12})(h_1^2 + \eta_1^2)h_2^2
\\
&& ~~~~~~
+ 2 |\lambda_1| \biggl( c_{\theta_1}
\left(h_2^2(h_1^2 - \eta_1^2) \right)
+ 2\, s_{\theta_1} h_2^2h_1\eta_1  \biggr)
 \biggr] \nonumber
\label{eq:tildeV-quartic}
\eea
where $\theta_1$ is the CP-violating phase of the $\lambda_1$ parameter.

%Considering the fact that only the inert fields are relevant during inflation, we can ignore the SM-like field, $h_3$ in Eq.~\eqref{eq:explicit-fields}, and study the inert 2-doublet system.

Further, we introduce another reparametrisation
\be
\label{eq:B1B2}
\eta_1 = \beta_1 \, h_1\,,
\qquad
h_2 = \beta_2 \, h_1 \, ,
\ee
with $\beta_1,\beta_2$ as field dependent values, to rewrite the potential as
\be
\label{eq:pot-with-B1B2}
\tilde{V} \approx
\frac{h_1^4}{4 \, \Omega^4} \biggl[
\lambda_{11}  (1+ \beta_1^2)^2
+\lambda_{22} \,  \beta_2^4
+ \biggl(
(\lambda_{12}+\lambda'_{12})(1 + \beta_1^2)
+ 2 |\lambda_1| \left( c_{\theta_1}
(1 - \beta_1^2)
+ 2\, s_{\theta_1} \beta_1 \right)\biggr) \beta_2^2
 \biggr]
\ee
Using this reparametrisation, one can also simplify the $\Omega^2$ parameter in Eq.~\eqref{eq:def-Omega} as
\be
\label{eq:def-OmegaSq}
\Omega^2 =
1+ \left(\frac{\xi_1}{M_{pl}^2}(1+\beta_1^2)+\frac{\xi_2}{M_{pl}^2}\beta_2^2+\frac{2 \,|\xi_4|}{M_{pl}^2}
\beta_2(c_{\theta_4} +\beta_1 s_{\theta_4})\right) \, h_1^2\equiv 1+\frac{B}{M_{pl}^2}h_1^2.
\ee
From Eq.~\eqref{eq:def-A}, recall that $\Omega^2= \exp(\tilde{A})$ using the shorthand notation
$\tilde{A}=\sqrt{\frac{2}{3}}\frac{A}{M_{pl}}$.
One can then write the field $h_1$ in terms of the reparametrised field $\tilde{A}$
%(or $\tilde{A}$ equivalently)
\be
\label{eq:h1-A-relation}
h_1^2 =
%\frac{M_{pl}^2}{2 \,|\xi_4|\,\beta_2(c_{\theta_4} + \beta_1\, s_{\theta_4}) } \,
\frac{M_{pl}^2}{B}
\left(e^{\tilde{A}} -1 \right)\,.
\ee
Therefore, expressing $h_1^2$ and $\Omega^2$ in terms of $\tilde{A}$ allows us to write the
potential in Eq.~\eqref{eq:pot-with-B1B2} in the form
\be
\tilde{V}\sim (1-e^{-\tilde{A}})^2 X(\beta_1,\beta_2).
\ee

We will be interested in the effect of the non-minimal coupling $\xi_4$ and
the associated phase $\theta_4$. Therefore, we will set $\xi_1=\xi_2=0$
and assume that the initial field values are such that $\Omega^2>0$ is guaranteed.
Therefore, with these assumptions,
the potential in Eq.~\eqref{eq:pot-with-B1B2} can be written as
\be
\label{eq:full-pot-B1B2}
\tilde{V}=
\biggl(\frac{M_{pl}^2}{2 \,|\xi_4|} \biggr)^2
\left(1-e^{-\tilde{A}}\right)^2 \,
X(\beta_1,\beta_2)\,
\ee
where
\begin{small}
\be
X(\beta_1,\beta_2) =
\frac{\lambda_{11}  (1+ \beta_1^2)^2
+\lambda_{22} \,  \beta_2^4
+ \left(
(\lambda_{12}+\lambda'_{12})(1 + \beta_1^2)
+ 2 |\lambda_1| \left( c_{\theta_1}
(1 - \beta_1^2)
+ 2\, s_{\theta_1} \beta_1 \right)\right) \beta_2^2 }
{4\beta_2^2(c_{\theta_4} + \beta_1\, s_{\theta_4})^2 } \,.
\ee
\end{small}
\begin{figure}[t!]
\begin{center}
\includegraphics[scale=0.7]{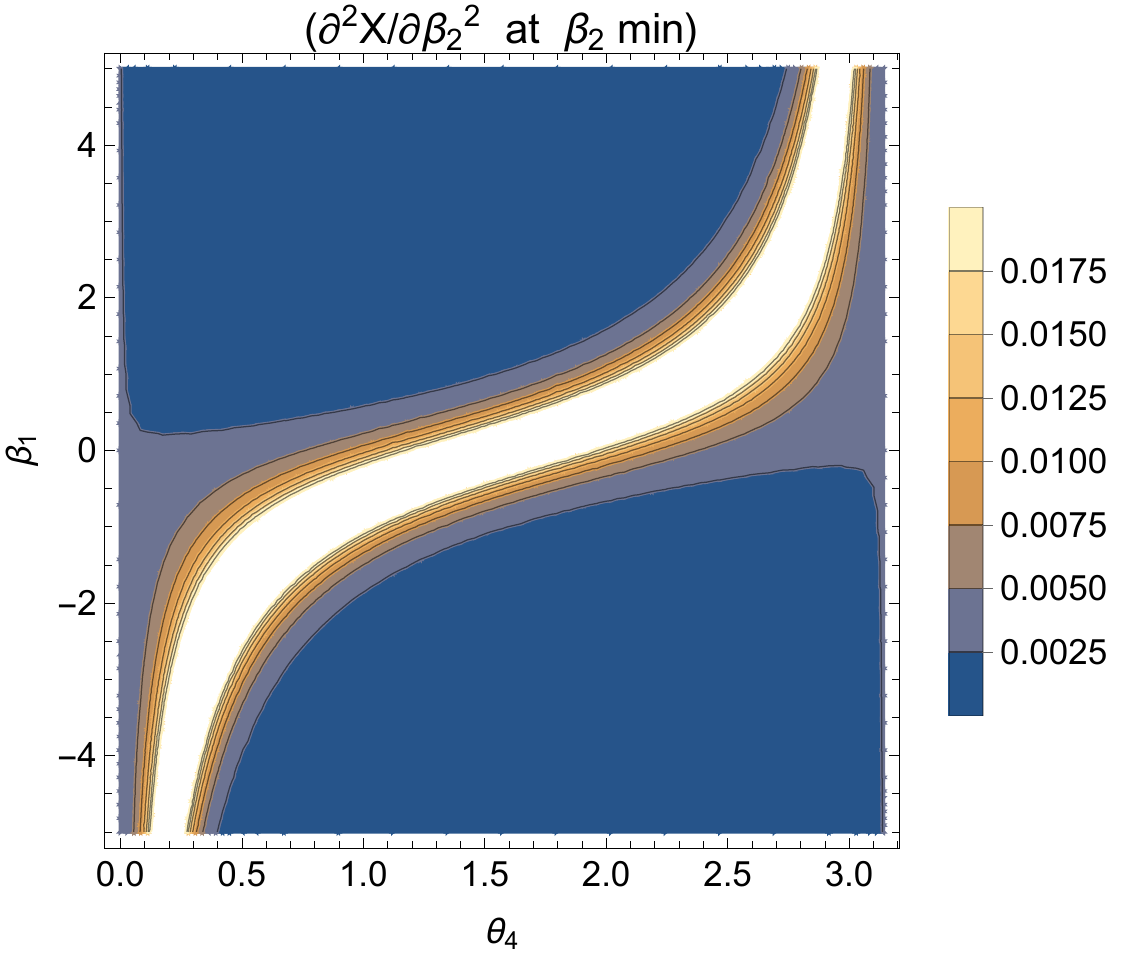}
\includegraphics[scale=0.7]{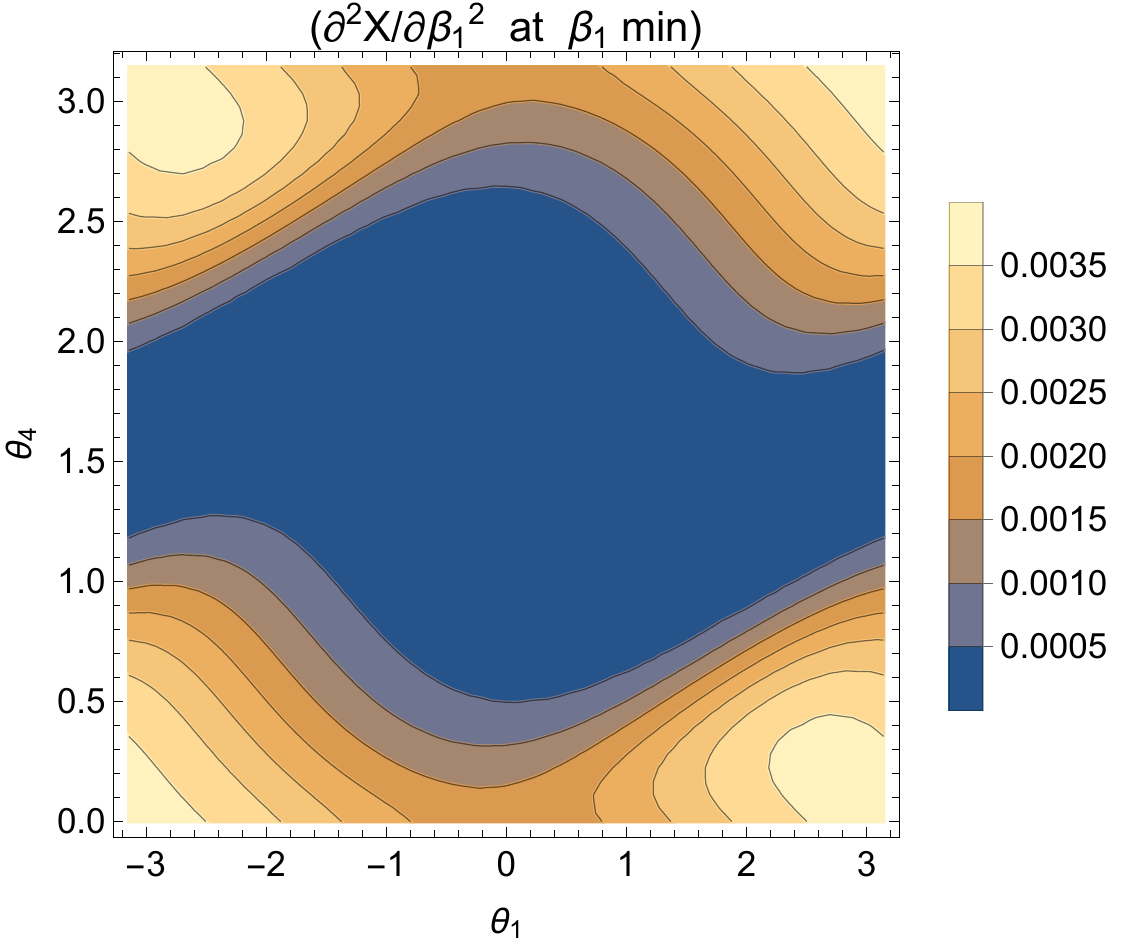}
\caption{The second order derivative of the function $X(\beta_1,\beta_2)$ with respect to $\beta_2$ at the minimum ($\partial X/\partial \beta_2=0$) on the left and the second order derivative of the function $X(\beta_1)$ with respect to $\beta_1$ at the minimum ($\partial X/\partial \beta_1=0$) on the right (all $\lambda_i \sim 0.001$). The white area on the left panel corresponds to where the denominator in Eq.~\eqref{eq:Xpp-B2} becomes zero.}
\label{Fig:Xpp-at-B2min}
\end{center}
\end{figure}
Following the procedure in \cite{Gong:2012ri}, to find the direction of inflation, we first minimise the $X(\beta_1,\beta_2)$ function with respect to $\beta_2$ which occurs at
\be
\frac{\partial X(\beta_1,\beta_2)}{\partial \beta_2}=0
\quad \Rightarrow \quad
\beta_2^2 = \sqrt{\frac{\lambda_{11}}{\lambda_{22}} } \, (1+\beta_1^2)
\label{eq:B2atXmin}
\ee
%\blue{
The second order derivative at this point is
\be
\frac{\partial^2 X(\beta_1,\beta_2)}{\partial \beta_2^{\;2}} =
\frac{2\, \lambda_{22}}{(c_{\theta_4}+\beta_1 s_{\theta_4})^2}
\label{eq:Xpp-B2}
\ee
which is always positive provided $\lambda_{22}>0$, as shown in the left panel in Figure \ref{Fig:Xpp-at-B2min}.

Using the $\beta_2$ value in Eq.~\eqref{eq:B2atXmin}, we can write the $X(\beta_1,\beta_2)$ function solely in terms of $\beta_1$,
\be
X(\beta_1)=
\frac{
(1+ \beta_1^2) \, \Lambda
+ 2  \left( (1-\beta_1^2) c_{\theta_1}
+ 2\beta_1 s_{\theta_1}\right) |\lambda_1|}{4\,(c_{\theta_4} + \beta_1\, s_{\theta_4})^2 }
\ee
with $\Lambda = \lambda_{12}+\lambda'_{12}+2\sqrt{\lambda_{11}\lambda_{22}}\,$.

We repeat the same treatment and minimise the $X(\beta_1)$ function with respect to $\beta_1$.
\be
\label{eq:B1atXmin}
\frac{\partial X(\beta_1)}{\partial \beta_1}=0
\quad \Rightarrow \quad
\beta_1 =
\frac{(\Lambda +2 |\lambda_1| c_{\theta_1}) s_{\theta_4} - 2 |\lambda_1| c_{\theta_4} s_{\theta_1}}{(\Lambda -2 |\lambda_1| c_{\theta_1}) c_{\theta_4} - 2 |\lambda_1| s_{\theta_4} s_{\theta_1}}
\ee
We check the positivity of the second order derivative at the minimum point which is satisfied for all $\theta_1,\theta_4$ values as shown in the right panel of Figure \ref{Fig:Xpp-at-B2min}.

\begin{figure}[t!]
\centering
\includegraphics[scale=0.7]{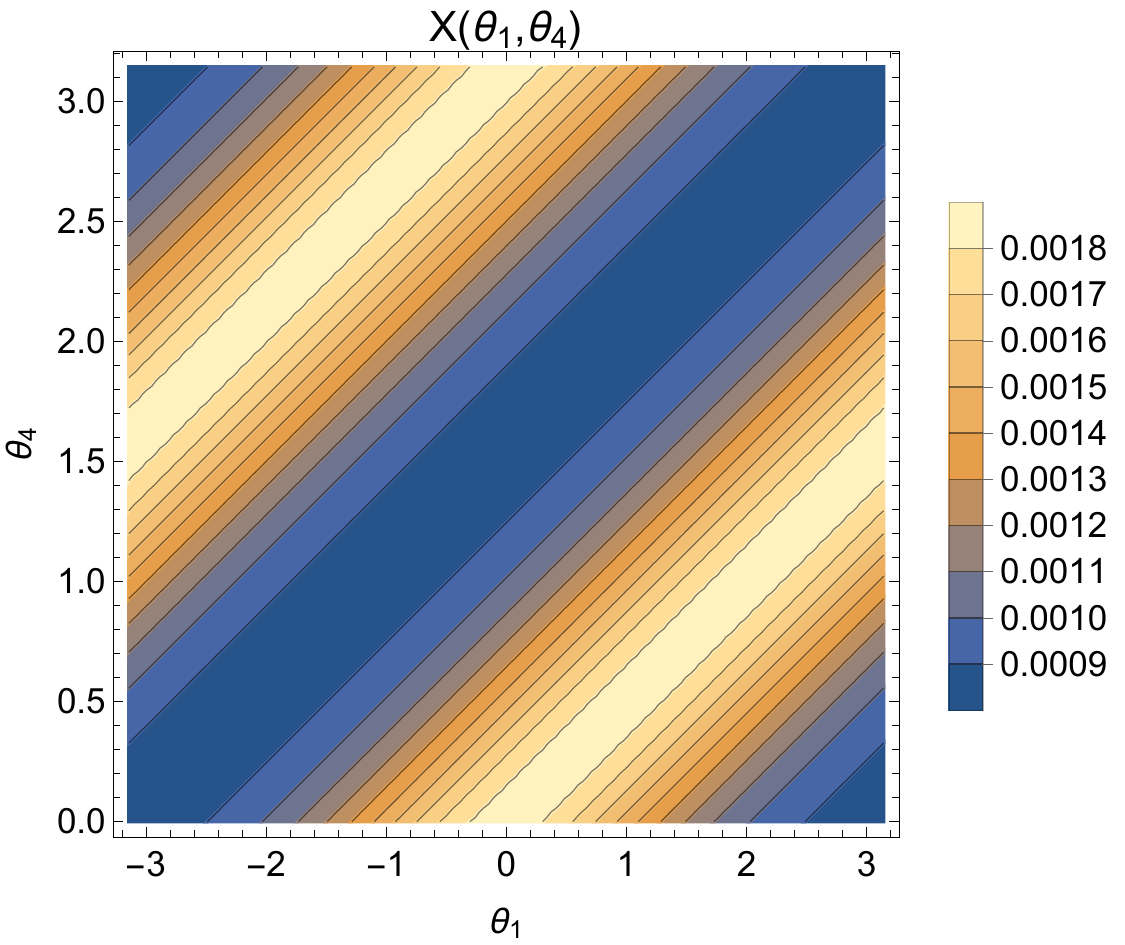}~
\includegraphics[scale=0.7]{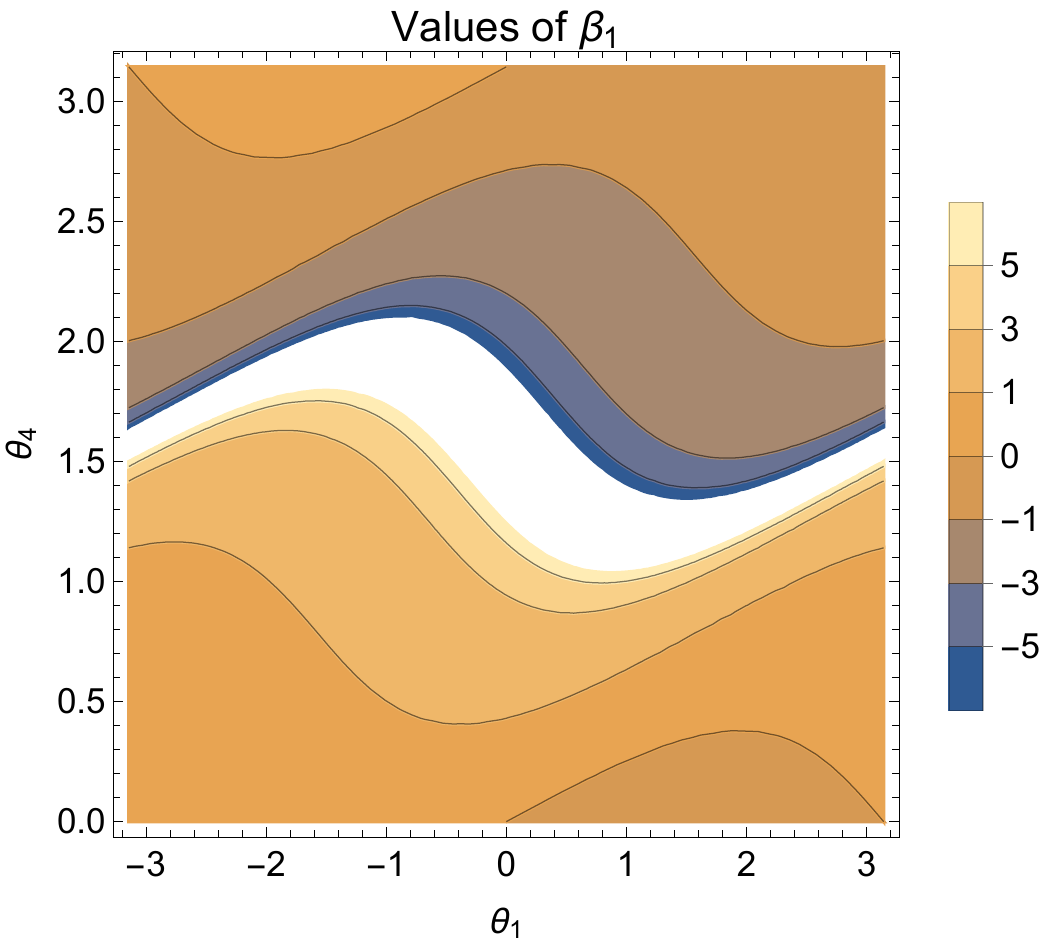}
%\\
%\includegraphics[scale=0.8]{3DPlot-X-T1-T4.pdf}~~
%\includegraphics[scale=0.75]{B1-3DPlot.pdf}
\caption{The $X(\theta_1,\theta_4)$ function on the left and the values of $\beta_1$ on the right for varying values of $\theta_1$ and $\theta_4$ (all $\lambda_i \sim 0.001$). The white region in the right panel shows a discontinuity where $\beta_1$ values tend to plus infinity approaching from the bottom and to minus infinity approaching from the top of the plot.}
\label{Fig:X-T1T4-B1B2}
\end{figure}

Replacing the $\beta_1$ value which minimises the $X(\beta_1)$ function back into the $X(\beta_1)$ function itself, yields the form of $X$ independent of $\beta_1$ and $\beta_2$ with only $\theta_1$ and $\theta_4$ as variables:
\be
\label{eq:X-thetas}
X(\theta_1,\theta_4) = \frac{\frac{1}{4}\Lambda^2 - \lambda_1^2  }{\Lambda - 2\lambda_1 \cos(\theta_1-2\theta_4)}
\ee
%}
The left panel in Figure \ref{Fig:X-T1T4-B1B2} shows the $X(\theta_1,\theta_4)$ function for allowed values of $\theta_1$ and $\theta_4$.
At each point in the plots, one can derive the values of $\beta_1$ and consequently $\beta_2$ using Eq.~\eqref{eq:B2atXmin} for given values of $\theta_1$ and $\theta_4$.
The right panel in Figure \ref{Fig:X-T1T4-B1B2} shows the values of $\beta_1$ for varying values of $\theta_1$ and $\theta_4$.

\section{Inflationary dynamics}
\label{sec:slow-roll}

\begin{figure}[t!]
\centering
\includegraphics[scale=0.8]{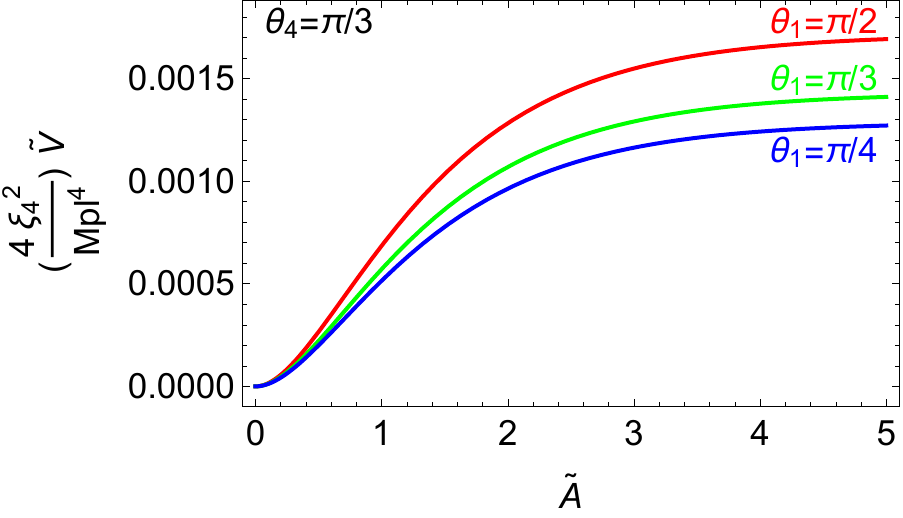}~~
\includegraphics[scale=0.8]{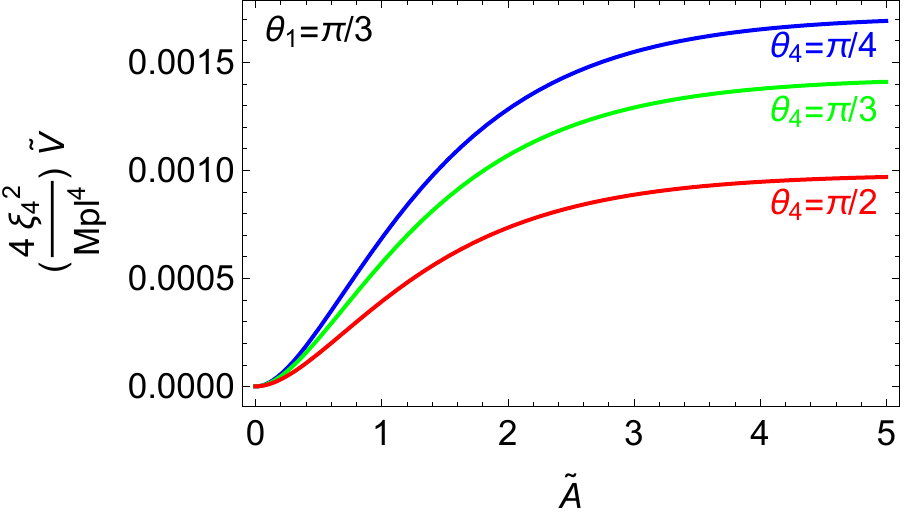}
\caption{The inflationary potential for different values of $\theta_1$ and $\theta_4$ (all $\lambda_i \sim 0.001$).}
\label{Fig:V-vs-ts}
\end{figure}

With the procedure used in the previous section,
%we have the function $X$ independent of $\beta_1,\beta_2$ and as a function of $\theta_1,\theta_4$.
the dynamics is essentially that of a single field inflation.
The full inflationary potential in Eq.~\eqref{eq:full-pot-B1B2} can be written as
\be
\label{eq:full-pot-T1T4}
\tilde{V}= \biggl(\frac{M_{pl}^2}{2 \,|\xi_4|} \biggr)^2
\left(1-e^{-\tilde{A}}\right)^2 \,
X(\theta_1,\theta_4)\,
\ee
Figure \ref{Fig:V-vs-ts} shows the inflationary potential for different values of  $\theta_1$ and $\theta_4$. Note that the potential is almost flat at high field values which ensures a slow roll inflation.

%\textbf{To calculate the slow roll parameters, the $X(\theta_1,\theta_4)$ function is irrelevant}
For the usual slow roll parameters in this case the function $X$ is irrelevant,
since it cancels in the expressions for $\epsilon$ and $\eta$, which are
\bea
\epsilon &=&
\frac{1}{2}M_{pl}^2 \,\left(\frac{1}{\tilde{V}}\frac{d\tilde{V}}{dA}\right)^2 =
\frac{4}{3\left( 1- e^{\tilde{A} }  \right)^2}\,,
\label{epsilon}
\\[3mm]
\eta &=&
M_{pl}^2 \, \frac{1}{\tilde{V}}\frac{d^2\,\tilde{V}}{dA^2}
=
\frac{4(2-e^{\tilde{A}})  }{3\left( 1- e^{\tilde{A} }  \right)^2}\,.
\label{eta}
\eea
For field values $A\gg M_{pl}$ (or equivalently $\tilde{A}\gg 1 $), both parameters $\epsilon,\eta \ll 1$ which satisfies the slow roll condition. Inflation ends when $\epsilon \simeq 1$.
To calculate the values of $A$ at the beginning and end of inflation,  $A_i$ and $A_f$ respectively, one needs to calculate the number of e-folds $N_e$, i.e. the number of times the universe expanded by $e$ times its own size. $N_e$ is calculated to be
\bea
\label{eq:efolds}
N_e = \frac{1}{M_{pl}^2}\int_{A_{f}}^{A_{i}} \frac{\tilde{V}}{\tilde{V}'}\,dA
\, = \,
\frac{3}{4}
\left[ \tilde{A}_f - \tilde{A}_i
-e^{\tilde{A}_f} + e^{\tilde{A}_i}
\right] ,
\eea
where $\tilde{V}'=\frac{d\tilde{V}}{dA}$ and $A_{i}$ ($\tilde{A}_i$) is the value of $A$ ($\tilde{A}$) at the beginning of inflation and $A_{f}$  ($\tilde{A}_f$) is the value of $A$ ($\tilde{A}$) at the end of the inflation.
Since inflation ends when $\epsilon\simeq 1$, one can calculate $A_{f}$, which yields:
\be
e^{\tilde{A}_f} = \exp\left(\sqrt{\frac{2}{3}}\frac{A_{f}}{M_{pl}}\right)  \simeq  2.1547
\quad \Rightarrow \quad
\tilde{A}_f = \sqrt{\frac{2}{3}}\,\frac{A_{f}}{M_{pl}} \simeq
0.7676 \, .
\label{eq:Af}
\ee
To calculate $A_{i}$, one could plug in the $A_{f}$ value into Eq.~\eqref{eq:efolds} assuming $N_e=60$, which results in
\be
\frac{3}{4}
\left[-\tilde{A}_i+  e^{\tilde{A}_i}
\right]- 1.0403
=60,
\quad
\Rightarrow  \quad
\tilde{A}_i = \sqrt{\frac{2}{3}}\frac{A_{i}}{M_{pl}}
\approx
4.4524
\ee

%%NEED TO DECIDE IF WE WANT TO KEEP THE FIRST PARAGRAPH HERE:
%% Well, it is referred to later, so let's keep it!

At this point we can also check
the field values in terms of the
original field $h_1$ using Eq.~\eqref{eq:h1-A-relation}.
% one can calculate the $h_1$ field values for $A_i$ and $A_f$, which are
This gives
\be
\label{eq:hi&hf}
 {h_1}_f  =
\frac{1.85 \times 10^{18}  }{\sqrt{|\xi_4|\,\beta_2(c_{\theta_4} + \beta_1\, s_{\theta_4}) }}\,,
\qquad
{h_1}_i  =
\frac{1.59 \times 10^{19} }{\sqrt{|\xi_4|\,\beta_2(c_{\theta_4} + \beta_1\, s_{\theta_4}) }} \,.
\ee
In the case of Higgs-inflation where the non-minimal coupling to gravity, $\xi$, is forced to be of the order $\sim 10^{4}$ GeV, %(to satisfy the $P_s$ bounds),
the $h$ field values during inflation are as large as $10^{16}$ GeV or so. In our case
the situation is similar.
%%UP TO HERE.

%%%%%%%%%%%%%%%%%%%%%%%%%%%%
% In our framework, the non-minimal coupling to gravity, $\xi_4$, can take any value from $\mathcal{O}( 10^{2} - 10^{5})$ GeV. Moreover, the pre-factor $1/\sqrt{\beta_2(c_{\theta_4} + \beta_1\, s_{\theta_4})}$ could take large values which leads to smaller $h_1$ values.
% For example, for large values of $\theta_4 \sim \pi/2$ and typical values of $\beta_1, \beta_2 \sim 100$, the $h_1$ values at the beginning of inflation could be as low as
% $\mathcal{O}( 10^{13} - 10^{15})$.
% {\bf{NOTE: We need to check that this is not tautologous in the sense that suppression by large
% $\beta_1$, $\beta_2$ implies that $\eta_1$ and $h_2$ are enhanced by the same factor
% relative to $h_1$!}} \blue{(It is tautologous. Maybe we should remove it then.)}
%

%In principle $N_e$ could be any number greater than around $50$ to solve flatness and horizon problems. $60$ e-folds solves the baryon asymmetry problem if inflationary energy scales are $O[10^{16}]\,\textrm{GeV}$ \cite{PhysRevD.78.123501}. Lower inflationary energy scales would need more e-folds and vice versa. However, the number of e-folds cannot be much larger than $60$.

\begin{figure}[t!]
\centering
\includegraphics[scale=0.6]{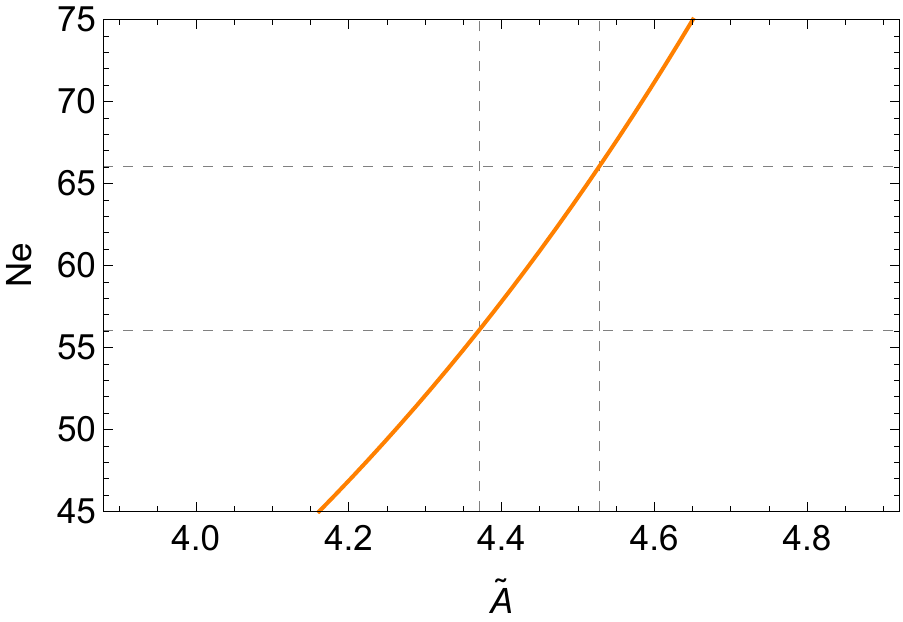}~
\includegraphics[scale=0.6]{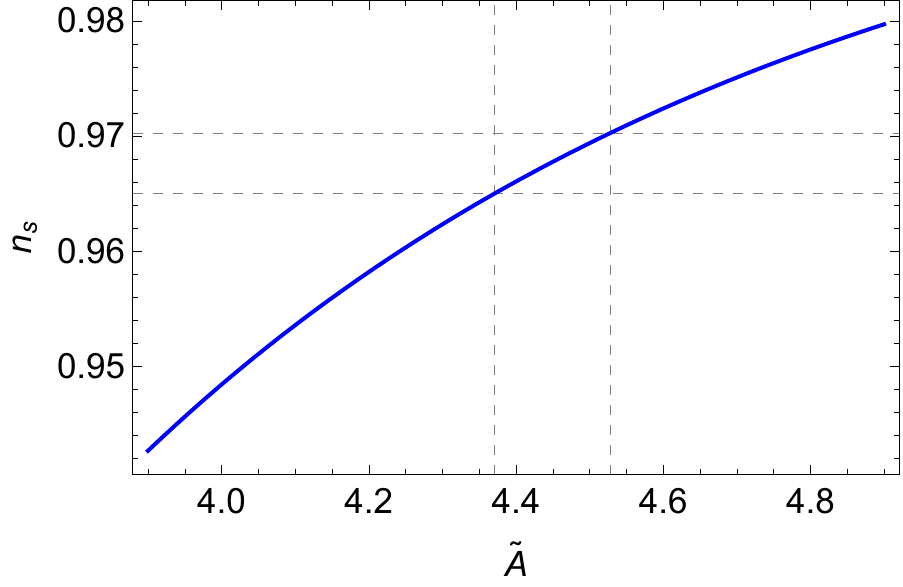}~
\includegraphics[scale=0.6]{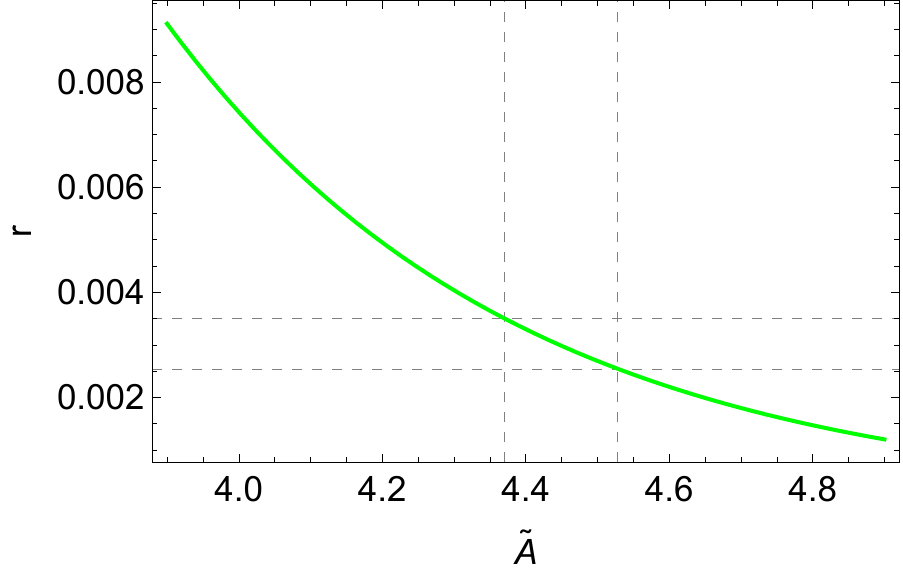}
\caption{The slow roll parameters: the number of $e$-folds $N_e$ (left), spectral index $n_s$ (center) and tensor to scalar ratio $r$ (right) as a function of $\tilde{A}$ with the grid-lines highlighting the $55 <N_e <65$ values.}
\label{Fig:Ps-r-ns}
\end{figure}

\begin{figure}[h!]
\centering
\includegraphics[scale=1]{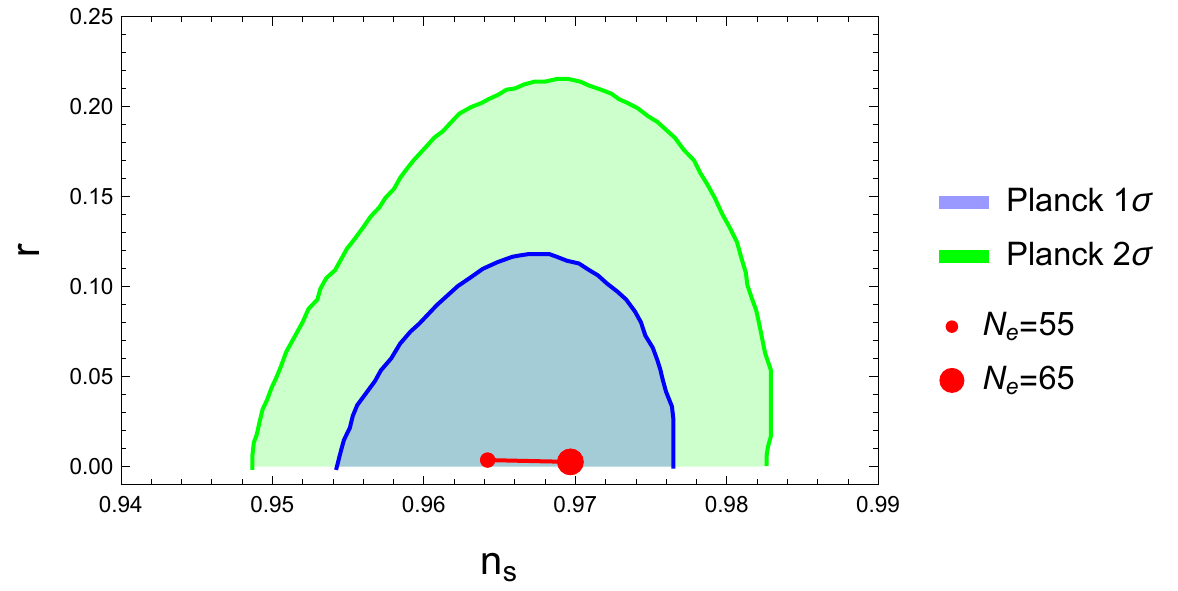}
\caption{The 1$\sigma$ and 2$\sigma$ regions for $n_s$ and $r$
from Planck observation compared to the theoretical
prediction of our framework.}
\label{Fig:Planck-nsVsr}
\end{figure}

Having fixed $N_e$ to 60, and calculated the $A$ field value at the start of inflation, we can derive the scalar power spectrum, $P_s$, the tensor to scalar ratio $r$ and the spectral index $n_s$ as follows:
\bea
P_s =\frac{1}{12\,\pi^2 \, M_{pl}^6} \frac{ \left({\tilde{V}}\right)^3}{\left({\tilde{V}'}\right)^2 }
&=&
\left(\frac{ (1-e^{\tilde{A}})^4}{128 \, \pi^2\, e^{2\tilde{A}}}\right)\frac{X(\theta_1,\theta_4)}{|\xi_4|^2}
=
5.565 \times \frac{X(\theta_1,\theta_4)}{|\xi_4|^2} ,
\\
\nonumber\\
r =16\,\epsilon &=& 0.00296,
 \\
n_s =1-6\epsilon+2\eta &=& 0.9678,
\eea
where $\tilde{V}'$ is the derivative of $\tilde{V}$ with respect to $A$ and both $\tilde{V}$ and $\tilde{V}'$ are calculated at the $A_{i}$.
Figure \ref{Fig:Ps-r-ns} shows the slow roll parameters $N_e$, $n_s$ and $r$ with respect to $\tilde{A}$ with the grid-lines highlighting the $55 <N_e <65$ values. We show the inflationary parameters over a range of $N_e$, since there is no reason for $N_e$ to be precisely $60$.
The values of $r$ and $n_s$ are well within the Plank bounds of $n_s=0.9677\pm 0.0060 $ at $1\sigma$ level and $r<0.11$ at $95\%$ confidence level \cite{Ade:2015lrj}. Note that the spectral index and the tensor to scalar ratio are in agreement with the Planck bounds over the full range of $N_e$.
Figure~\ref{Fig:Planck-nsVsr} shows the 1$\sigma$ and 2$\sigma$ regions allowed by Planck observations in the $r$-$n_s$ plane and   the theoretical predictions of our framework for $N_e$ values of 55 and 65.

\begin{figure}[t!]
\centering
\includegraphics[scale=0.8]{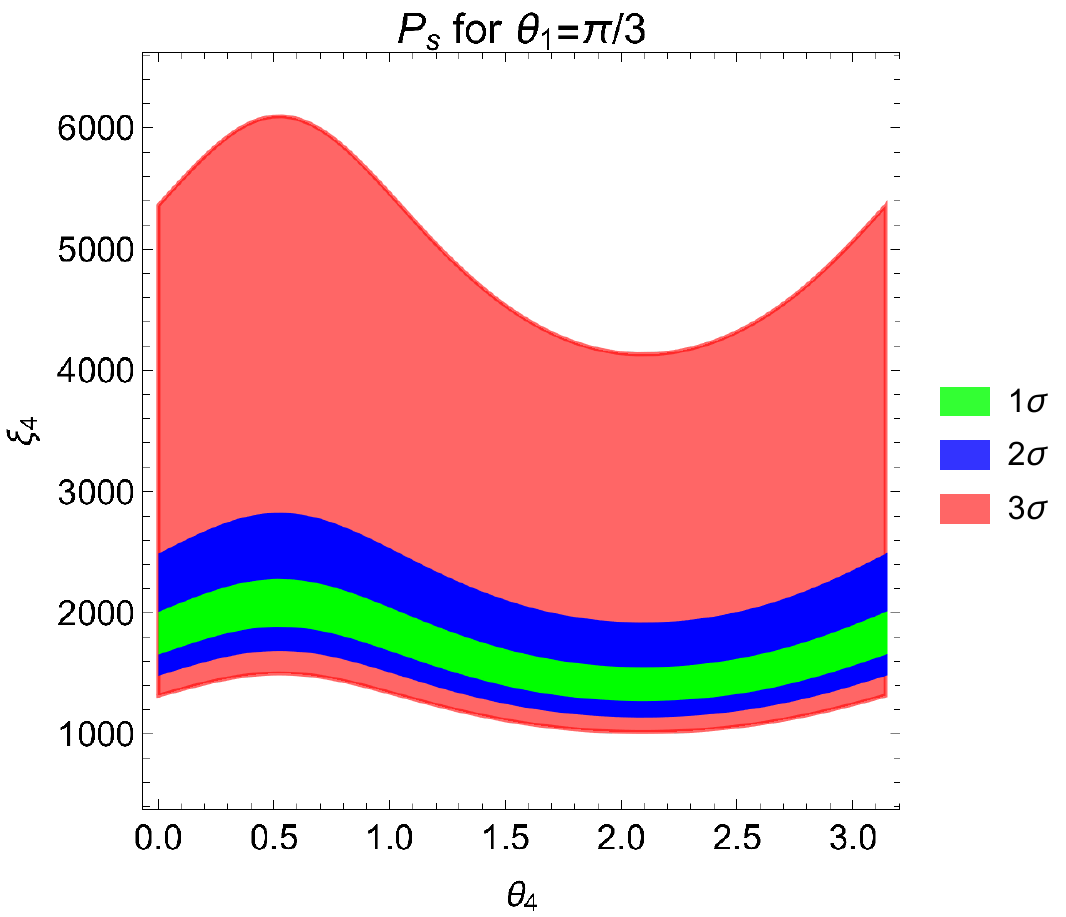}~
\includegraphics[scale=0.75]{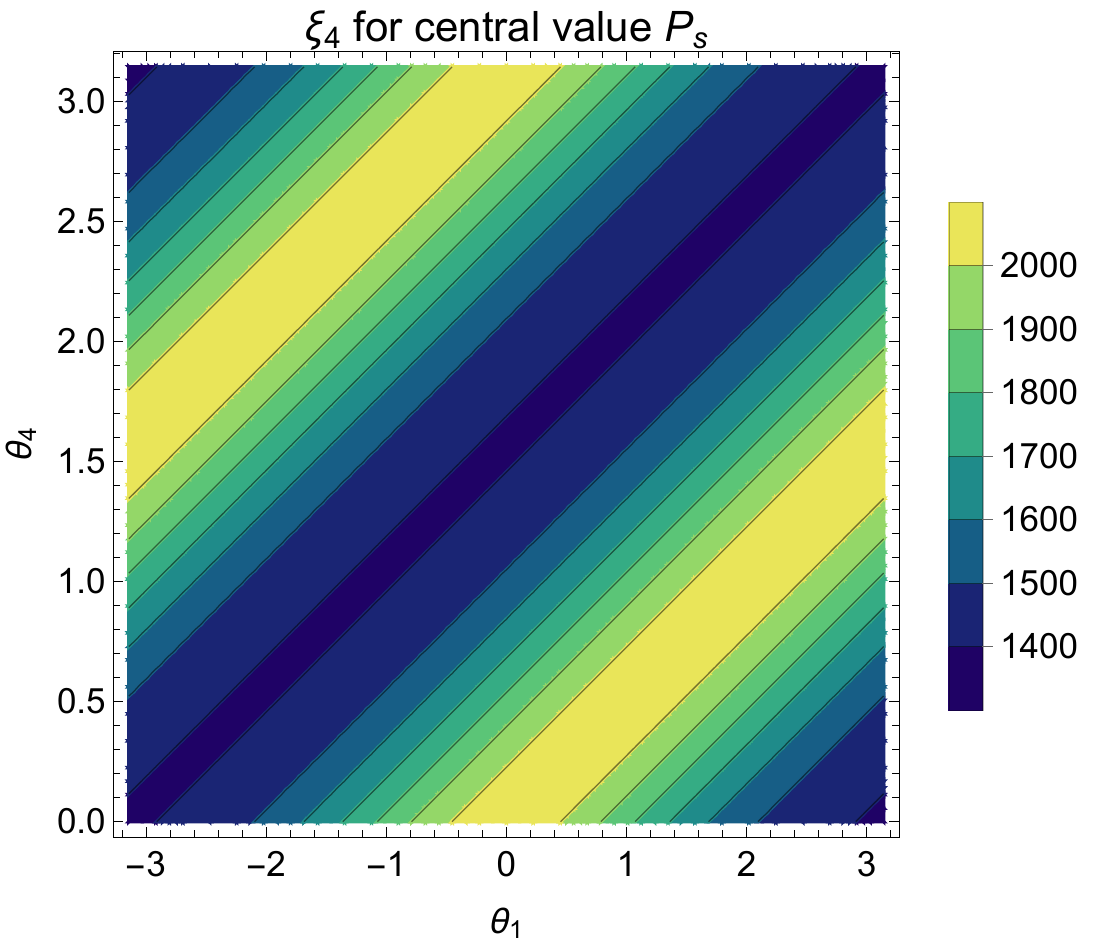}
%\\
%\includegraphics[scale=0.8]{Ps-t1-t4-3DPlot.pdf}~~
%\includegraphics[scale=0.75]{xi4-at-PsCV-3D.pdf}
\caption{Left panel: $P_s$ values for the fixed $\theta_1=\pi/3$ angle and varying values of $\xi_4$ and $\theta_4$ up to $3\sigma$ standard deviation from the observed central value. Right panel: Contours of $\xi_4$ in the $\theta_1$-$\theta_4$ plane which lead to $P_s$ central values (all $\lambda_i \sim 0.001$).}
\label{Fig:Ps-Xi4}
\end{figure}

Observations from WMAP7 \cite{Komatsu:2010fb} constrain the scalar power spectrum which put a bound on the $\xi_4$  coupling and angles $\theta_1,\theta_4$,
\be
\label{eq:PS-bound}
P_s=(2.430\pm 0.091)\times 10^{-9}
\, = \,
5.565 \times \frac{X(\theta_1,\theta_4)}{|\xi_4|^2} \, .
\ee
In the left panel of Figure \ref{Fig:Ps-Xi4}, we show $P_s$ values for the fixed $\theta_1=\pi/3$ angle and varying values of $\xi_4$ and $\theta_4$ up to $3\sigma$ standard deviation from the central value in Eq.~\eqref{eq:PS-bound}. In the right panel, we fix $P_s$ to the WMAP7 central value for fixed values of $\lambda_i \sim 0.001$ to get
\be
%|\xi_4|^2 = 2.29 \times 10^{9}\, X(\theta_1,\theta_4)
%\qquad
%\Rightarrow
%\qquad
|\xi_4|= 4.785\times 10^{4}\, \sqrt{X(\theta_1,\theta_4) }\,
\ee
and show contours of $\xi_4$ for varying values of $\theta_1$ and $\theta_4$.
Note that every point in the plot yields the exact $P_s$ central value.

This is a very important feature of our framework. To satisfy the bounds on the scalar power spectrum, the function $X(\theta_1,\theta_4)$ allows for a wide range of $\xi_4$ values as shown in Figure~\ref{Fig:Ps-Xi4}. This is in contrast to
the Higgs-inflation models where $P_s \propto {\lambda}/{\xi^2}$
with $\lambda$ the Higgs self-coupling which is fixed to be $\sim 0.12$ at the electroweak scale. Thus, for $P_s$ to agree with observations at the inflationary scale, $\xi$ will have to be very large $\mathcal{O}(10^4)$.
In our set-up, a combination of parameters $\lambda_{1}, \lambda_{11}, \lambda_{22}, \lambda_{12},\lambda'_{12}$ appears in the $X(\theta_1,\theta_4)$ function. The only constraint limiting these parameters is the stability of the potential requiring
\bea
\lambda_{ii} > 0, \qquad
\lambda_{ij} + \lambda'_{ij} > -2 \sqrt{\lambda_{ii}\lambda_{jj}}, \qquad
|\lambda_i| \leq |\lambda_{ii}|, |\lambda_{ij}|, |\lambda'_{ij}| , \quad i\neq j = 1,2,3\, ,
\eea
which allows for very small values of $\lambda_i \sim 0.001$ which, in turn, allows for much smaller values of $\xi_4$, at least one order of magnitude than the $\xi$ value in Higgs-inflation models.

\section{Reheating and scalar asymmetries}
\label{sec:reheating}

At the end of inflation, the energy stored in the inflaton disperses as the inflaton decays/annihilates into the SM particles through processes mediated by the SM-Higgs and gauge bosons in our case, during the so-called reheating phase \cite{Linde:1981mu}. There are numerous
details on how the inflaton decays and creates the initial condition for the
conventional hot early universe. Here our main interest is to discuss how the CP asymmetry
originating from the non-minimal coupling, is transferred to the SM
degrees of freedom.

For the discussion of the scalar asymmetries, let's focus on the neutral components of the $\phi_1$ doublets acquiring an initial non-vanishing expectation value at the exit from
inflaton. We write the field fluctuations around the initial conditions as
\be
\label{eq:fluctuations}
\left\{\begin{array}{c}
\phi_1 \to \phi_1 - a_1 e^{i \,\alpha} \, ,\quad
\phi_1^\dagger \to \phi_1^* - a_1 e^{-i \,\alpha}
\\[2mm]
\phi_2 \to \phi_2 - a_2 \, , \phantom{e^{i \,\alpha}} \quad
\phi_2^\dagger \to \phi_2^* - a_2  \phantom{e^{-i \,\alpha}}
\\[2mm]
\phi_3 \to \phi_3 - a_3 \, , \phantom{e^{i \,\alpha}}\quad
\phi_3^\dagger \to \phi_3^* - a_3  \phantom{e^{-i \,\alpha}}
\end{array}
\right.
\ee
The phase $\alpha$ here is related to the CP-violating phases of inflation.
Note that at the end of inflation the $h_1$ field has taken the value $h_{1f}$ according to Eq.~\eqref{eq:hi&hf} which is dependant on the inflationary dynamics, namely $\theta_4$, $\beta_1$ and $\beta_2$ which are dependant on $\theta_1$.
Since $h_1$ is the real part of the complex field $\phi_1$, its value is what feeds the
$a_1 \cos\alpha$ component of fluctuations in
Eq.~\eqref{eq:fluctuations}. The imaginary part of $\phi_1$, represented by
$\eta_1$, takes a value proportional to $\eta_{1f}=\beta_1 h_{1f}$ as shown in
Eq.~\eqref{eq:B1B2}, and feeds the $a_1 \sin\alpha$ component of the field fluctuations.
Recall that one can obtain the values of $\beta_1$ and $\beta_2$ for any given value of $\theta_1$ and $\theta_4$ from Eq.~\eqref{eq:B2atXmin} and Eq.~\eqref{eq:B1atXmin}.
Explicitly, one can write
\be 
\tan\alpha =\frac{a_1 \sin\alpha}{a_1 \cos\alpha} = \frac{\eta_{1f}}{h_{1f}} =   \beta_1 =
\frac{(\Lambda +2 |\lambda_1| c_{\theta_1}) s_{\theta_4} - 2 |\lambda_1| c_{\theta_4} s_{\theta_1}}{(\Lambda -2 |\lambda_1| c_{\theta_1}) c_{\theta_4} - 2 |\lambda_1| s_{\theta_4} s_{\theta_1}}\,,
\ee
with $\Lambda = \lambda_{12}+\lambda'_{12}+2\sqrt{\lambda_{11}\lambda_{22}}\,$ as mentioned before.
However, to keep the present discussion more transparent, we retain
a generic phase $\alpha$ here.

To discuss the consequences of this complex phase, we now
assume instant reheating. Since the field $\phi_3$ is light with respect to the inflaton
degrees of freedom, we expect the latter to quickly decay to
$\phi_3$. The asymmetry arising from the values of the fields
in Eq.~\eqref{eq:fluctuations} will manifest in creation of unequal number
of $\phi_3$ and $\phi_3^\ast$ quanta as follows.

Let us study the decay process $\phi_1 \to \phi_3^* \phi_3^*$ in detail.
From the potential in Eq.~\eqref{eq:V0-3HDM}, the amplitude of the tree-level process is proportional to
\bea 
\mathcal{M}_{(\phi_1 \to \phi_3^* \phi_3^*)} \; \propto \; - 2 a_1 \lambda_3 \, e^{i (\alpha +\theta_3)}
\qquad &\mbox{and}& \qquad 
\mathcal{M}_{(\phi_1^* \to \phi_3 \phi_3)} \; \propto \; - 2 a_1 \lambda_3 \, e^{-i (\alpha +\theta_3)}\,.
\eea
The generation of the asymmetry is sensitive to the interference between the tree and loop diagrams \cite{Balaji:2004xy,Balaji:2005ha}. Hence, we need to sketch what happens at loop level.
At one loop level, there are many diagrams that contribute to this decay process. 
For the purpose of demonstration, we consider the bubble diagrams which convert $\phi_1$ to $\phi_3$ with only $\phi_1$ and $\phi_1^{*}$ in the loop, as shown in Figure \ref{fig:tree-loop}. 
Clearly one needs to take into account all diagrams contributing to this decay process, specially since there may be interferences cancelling the CP asymmetry. 
However, since all triple scalar couplings in the potential can be different, one can ensure that such cancellation does not occur.
More careful analysis of these effects is deferred to a future work. 

\begin{minipage}{\linewidth}
\begin{figure}[H]
\begin{center}
\begin{tikzpicture}[thick,scale=1.0]
%\fill[black] (1.5,0) circle (0.06cm);
\draw (1.5,-1.25) -- node[black,above,xshift=-0.1cm,yshift=0.0cm] {} (1.5,-1.25);
\draw[dashed] (0,0) -- node[black,above,xshift=-1cm,yshift=-0.25cm] {$\phi_1$} (1.5,0);
\draw[dashed] (1.5,0) -- node[black,above,xshift=0.8cm,yshift=0cm] {$\phi_3^*$} (2.5,0.6);
\draw[dashed] (1.5,0) -- node[black,above,yshift=-0.6cm,xshift=0.8cm] {$\phi_3^*$} (2.5,-0.6);
\end{tikzpicture}
\hspace{15mm}
\begin{tikzpicture}[thick,scale=1.0]
%\fill[black] (1,0) circle (0.06cm);
\draw[dashed] (0,0) -- node[black,above,sloped,yshift=-0.25cm,xshift=-0.8cm] {$\phi_1$} (1,0);
\draw[dashed]  (1,0) node[black,above,sloped,yshift=-1.25cm,xshift=0.6cm] {$\phi_1^*$}  arc (180:0:0.6cm) ;
\draw[dashed]  (1,0) node[black,above,sloped,yshift=0.55cm,xshift=0.6cm] {$\phi_1$}  arc (-180:0:0.6cm) ;
%\fill[black] (2.2,0) circle (0.06cm);
\draw[dashed] (2.2,0) -- node[black,above,sloped,yshift=0cm,xshift=0cm] {$\phi_3$} (3.2,0);
%\fill[black] (3.2,0) circle (0.06cm);
\draw[dashed] (3.2,0) -- node[black,above,xshift=0.8cm,yshift=0cm] {$\phi_3^*$} (4.2,0.6);
\draw[dashed] (3.2,0) -- node[black,above,yshift=-0.6cm,xshift=0.8cm] {$\phi_3^*$} (4.2,-0.6);
\end{tikzpicture}
\caption{The tree level decay process $\phi_1 \to \phi_3^* \phi_3^*$ and the one-loop bubble diagram with $\phi_1$ and $\phi_1^*$ in the loop.}
\label{fig:tree-loop} 
\end{center}
\end{figure}
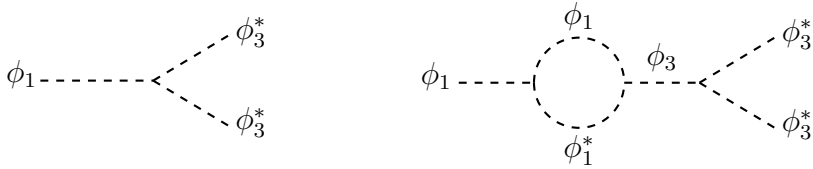    
\end{minipage}

The amplitude of the loop process with $\phi_1$ and $\phi_1^*$ running in the loop is proportional to
\bea 
\mathcal{M}_{(\phi_1 \to \phi_3 \to \phi_3^* \phi_3^*)}  &\propto & 
-4 a_1 a_3^2 \lambda_{11}\lambda_{33} (\lambda_{31}+\lambda'_{31}) e^{-i\alpha}\,,
\\
\mathcal{M}_{(\phi_1^* \to \phi_3^* \to \phi_3 \phi_3)}  &\propto & 
-4 a_1 a_3^2 \lambda_{11}\lambda_{33}
(\lambda_{31}+\lambda'_{31}) e^{i\alpha}\,.
\eea  
Due to the interference of the tree and loop diagrams, the decay processes are CP-violating and result in unequal number of $\phi_3$ and $\phi_3^*$ states. 
Consequently, we define the asymmetry $A^1_{CP}$ as the difference between the $\phi_1$ decay rate and its conjugate, and we find
\be
A^1_{CP} = \Gamma^{\mathrm{tree}+\mathrm{loop}}_{(\phi_1 \to \phi_3^* \phi_3^*)} -  \Gamma^{\mathrm{tree}+\mathrm{loop}}_{(\phi_1^* \to \phi_3 \phi_3)}   
\;=\; -\frac{1}{16\sqrt{3}\pi^2} a_1^2 a_3^2 \lambda_3 \lambda_{11}\lambda_{33} (\lambda_{31}+\lambda'_{31})\sin(2\alpha+\theta_3)\,.
\ee

This asymmetry in the scalar sector is then transferred to the fermion sector through the couplings of the Higgs field (the $\phi_3$ doublet) with the fermions.
For example, assuming the existence of right-handed neutrinos, the Yukawa interactions
between neutrinos and $\phi_3$ will generate an asymmetry between $\nu_L$ and $\bar{\nu}_R$,
which would be further translated into baryon asymmetry by the electroweak sphalerons.

\section{Conclusion and outlook}
\label{sec:conclusions}
Scalar fields which have non-minimal couplings to gravity are well-motivated inflaton candidates. Paradigmatic examples are the Higgs-inflation \cite{Bezrukov:2007ep}
and $s$-inflation models \cite{Enqvist:2014zqa}.
In this paper we have considered a scenario where several non-minimally coupled scalars
contribute to the inflationary dynamics. In particular we investigated a model where
these scalars are electroweak doublets and therefore generalize the Higgs inflation.
We focused on a setting where the dominant non-minimal coupling is allowed to be complex
and investigated the effect that this would have on CP-violation in our universe.
We determined the inflationary dynamics in the regime where the model essentially
conforms to the predictions of single field inflation. The essential difference is
that the inflaton obtains a non-zero phase representing possible source of CP-violation
for subsequent post-inflationary evolution.
At the end of inflation, the inflaton particle which is naturally assumed to have couplings with the SM Higgs, dumps its energy into the SM particle bath through the process of
reheating, which populates the universe with the SM particles. We sketched how
the complex value of the inflaton field leads to an asymmetry in the scalar sector decays,
and how this asymmetry will further be transmitted to the fermion sector.
There are numerous details in our scenario which can be investigated in more detail.
These include the multi-field dynamics during the inflation as well as the details
of reheating and subsequent particle decays. Also the detailed analysis of the
effects on the generation of baryon asymmetry need to be addressed in more detail.
We will consider these in future work on the model introduced in this paper.

\subsubsection*{Acknowledgements}
VK acknowledges financial support from Academy of Finland projects ``Particle cosmology and gravitational waves'' No. 320123
and ``Particle cosmology beyond the Standard Model'' No. 310130.

\end{document}